\documentclass[%
 reprint,
 amsmath,amssymb,
 aps,prl,
]{revtex4-2}

\usepackage{graphicx}
\usepackage{dcolumn}
\usepackage{bm}
\usepackage{xcolor}
\usepackage{soul}
\usepackage{upgreek}

\providecommand{\abs}[1]{\lvert#1\rvert}

\begin{document}

\preprint{APS/123-QED}

\title{Stagnation points control chaotic fluctuations in viscoelastic porous media flow}

\author{Simon J. Haward}
\author{Cameron C. Hopkins}
\author{Amy Q. Shen}
\affiliation{Okinawa Institute of Science and Technology Graduate University, Onna-son, Okinawa, 904-0495, Japan
}%

\date{\today}

\begin{abstract}

Viscoelastic flows through porous media become unstable and chaotic beyond critical flow conditions, impacting industrial and biological processes. Recently, Walkama \textit{et al.} [Phys. Rev. Lett. \textbf{124}, 164501 (2020)] have shown that geometric disorder greatly suppresses such chaotic dynamics. We demonstrate experimentally that geometric disorder \textit{per se} is not the reason for this suppression, and that disorder can also promote choatic fluctuations, given a slightly modified initial condition. The results are explained by the effect of disorder on the occurrence of stagnation points exposed to the flow field, which depends on the initially ordered geometric configuration. 
\end{abstract}

\maketitle

Unlike viscous Newtonian liquids (e.g., water), many fluids exhibit an elastic resonse to an applied strain. Such ``viscoelastic" fluids are widespread in biology (blood, mucus, synovial fluid) and industry (paints, coatings, fracking fluids). The elasticity is imparted by a mesoscopic structure (formed by e.g., polymers, proteins, or self-assemblies of lipids or surfactants), that relaxes after deformation \cite{Larson1999}. The strength of the elastic response is quantified by the Weissenberg number $\text{Wi} = \tau \dot\gamma$, with $\tau$ the fluid relaxation time and $\dot\gamma$ the rate of strain. While flows of Newtonian fluids become unstable and turbulent due to the onset of inertial effects at high Reynolds number, $\text{Re}\gg1$, viscoelastic flows can become unstable and ``elastically turbulent" even for $\text{Re} \ll 1$, purely due to elastic effects at high $\text{Wi}$  \cite{Larson1990,McKinley1991,Shaqfeh1996,Groisman2000,Zilz2012,Steinberg2021}. 

Viscoelastic porous media flow occurs in diverse processes from enhanced oil recovery (EOR) and filtration to drug delivery \cite{Anbari2018,Eberhard2020}. Porous media flow subjects a fluid to a complex cycle of deformation with high shear rates through the pore-throats or between obstacles, and high elongational rates at points of constriction or at stagnation points. Stagnation points are regions where particularly high fluid strains develop, resulting in strong stretching of the microstructure if $\text{Wi} \gtrsim 1$ \cite{DeGennes1974,Perkins1997,Francois2008,Haward2012,Kawale2017b,Haward2019}. Elastic tensile stresses due to stretching on curvilinear streamlines (as through porous media) are conditions well-established to lead to linear instabilities in viscoelastic fluids \cite{Pakdel1996,McKinley1996,Morozov2007,Muller2008}, which can be precursors to elastic turbulence \cite{Pan2013,Qin2017,Varshney2017,Sousa2018,Steinberg2021}. Resulting chaotic fluctuations are expected to greatly enhance the pressure loss and the dispersion in porous media, with positive impacts for e.g., removing oil ganglia from the pore space in EOR \cite{De2017,Ekanem2020,Browne2020b}.

\begin{figure}
    \centering
    \includegraphics[width=8.6cm]{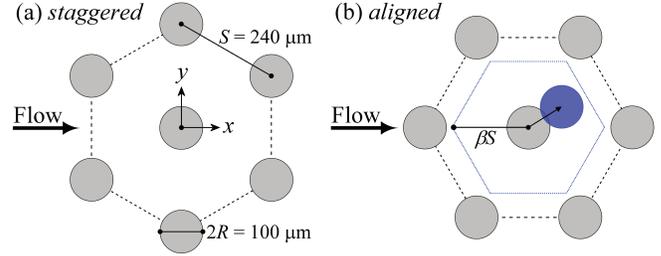}
\vspace{-0.25in}
    \caption{Unit cell representations of two contrasting ordered hexagonal arrays of posts used in the flow experiments. In (a) the posts are ``staggered" along the $x$-direction in which the flow is imposed. The post radius is $R$, lattice spacing $S$. Rotating the array by 30$^{\circ}$ aligns the posts in the flow direction (b). Disordered ``aligned" arrays are generated by the random displacement of each post within a hexagon of circumradius $\beta S$, as described in Ref.~\onlinecite{Walkama2020}. 
 }
    \label{schemes}
\vspace{-0.05in}
\end{figure}

Recently, in microfluidic experiments with polymer solutions in porous media modeled by hexagonal arrangements of posts, Walkama \textit{et al.} demonstrated that introducing random disorder to the array can significantly suppress chaotic fluctuations for $\text{Wi} \gtrsim 1$ \cite{Walkama2020}. Disorder resulted in the opening of clear flow paths for the viscoelastic fluid, which became more dominated by shear and less by extensional kinematics as disorder increased, thus reducing the stretching of the dissolved polymer, and suppressing fluctuations. However, only one initially ordered configuration of posts [similar to Fig.~\ref{schemes}(a)] was considered. Other works show that instabilities and fluctuations in viscoelastic flows through ordered post arrays depend on the orientation of the array relative to the flow direction \cite{Kawale2017}. Different behavior might be expected from an array with posts staggered along the flow direction [Fig.~\ref{schemes}(a)] than from the same array rotated by 30$^{\circ}$ such that the posts are aligned [Fig.~\ref{schemes}(b)].

\begin{figure*}[ht!]
    \centering
    \includegraphics[width=17.2cm]{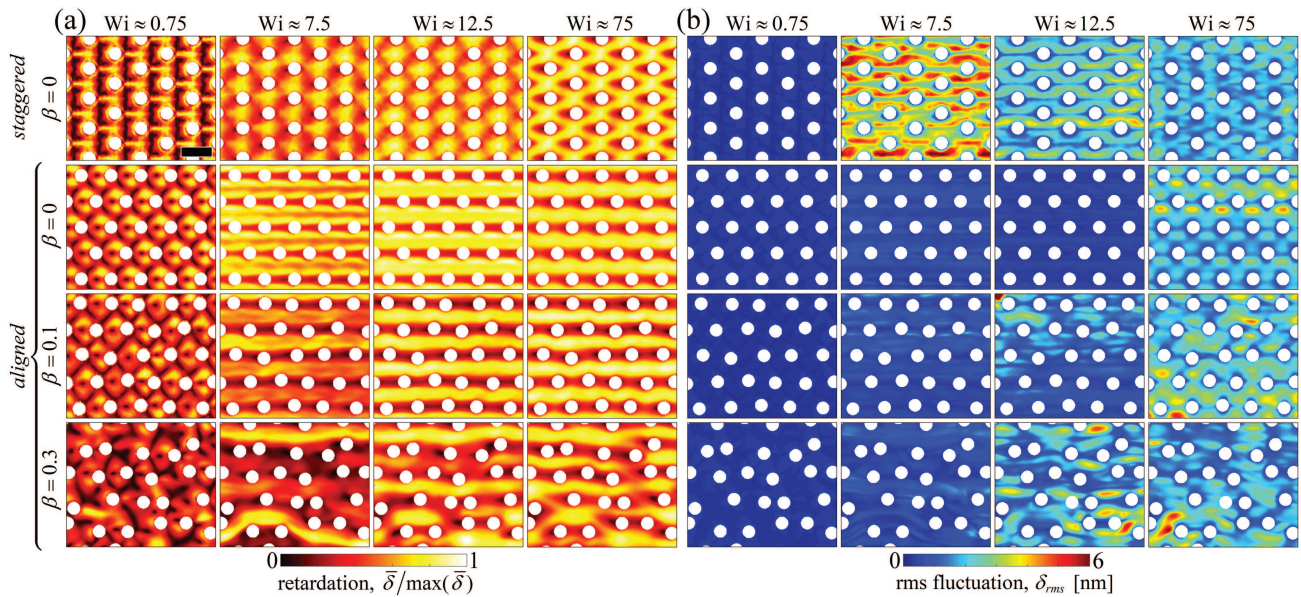}
\vspace{-0.1in}
    \caption{Evolution of the retardation fields with increasing $\text{Wi}$ for the WLM solution in post arrays with various $\beta$: (a) normalized, time-averaged retardation fields $\bar{\delta}/\max(\bar{\delta})$; (b) local retardation fluctuations $\delta_{rms}$.  Field of view is centered on the coordinate origin at the center of each array (54\% of full field of view shown). Scale bar: $250~\upmu$m.
}
    \label{images}
\vspace{-0.05in}
\end{figure*}

In this Letter, we present microfluidic experiments of viscoelastic flow through two ordered arrays of posts oriented as shown in Fig.~\ref{schemes}. First, we demonstrate that the chaotic fluctuations at $\text{Wi} \gtrsim 1$ can be strongly supressed in a perfectly ordered array simply by aligning the posts in the flow direction [Fig.~\ref{schemes}(b)]. Subsequently disordering the aligned array does not suppress, but \emph{promotes} chaotic fluctuations over a wide range of $\text{Wi}$. Our results are explained (in context with those of Walkama \textit{et al.}  \cite{Walkama2020}) by the presence and role of stagnation points in the flow field, as opposed to the degree of disorder \textit{per se}.

Microfluidic channels ($W = 2.4$~mm wide, $H=1$~mm high, and 25~mm long) containing arrays of $\approx 300$ circular posts (radius $R=50~\upmu$m) were fabricated in fused silica by selective laser-induced etching~\cite{Burshtein2019}. Two of the channels contained ordered hexagonal arrays (disorder $\beta = 0$, lattice spacing $S=240~\upmu$m, porosity $\phi \approx 0.84$) in either ``staggered" or ``aligned" orientations [Fig.~\ref{schemes}]. 
Five arrays were generated by disordering the aligned configuration with $\beta = [0.05 ,~0.1,~0.2,~0.3,~0.4]$ (Fig.~\ref{schemes}(b) \cite{Walkama2020}).

The model viscoelastic test fluid is an aqueous wormlike micellar (WLM) solution of 100 mM cetylpyridinium chloride and 60 mM sodium salicylate \cite{Rehage1991}. The fluid is shear-thinning with a zero shear viscosity $\eta_{0} \approx 48$~Pa~s and a single-mode Maxwell relaxation time $\tau = 1.5$~s. Flow through the post arrays is driven by a syringe pump (Cetoni GmbH) infusing at controlled volumetric flow rate $Q$, hence average flow velocity ${U=Q/\phi WH}$, and characteristic deformation rate $\dot\gamma = U/R$. The maximum Reynolds number, $\text{Re} = \rho U R/\eta(\dot\gamma) < 10^{-5}$ (density $\rho$, viscosity $\eta(\dot\gamma)$), means that inertia can be neglected.

A high speed polarizing camera (Photron CRYSTA PI-1P, see Ref.~\onlinecite{Hopkins2020}) is used to visualize the regions of high micelle orientation and elastic stress in the post arrays over a range of imposed $\text{Wi}=\tau \dot\gamma$, via the flow-induced optical retardation, $\delta$. Time averaged fields $\bar\delta = \langle\delta\rangle_t$, captured at 125~Hz are shown for a few of the arrays and several values of $\text{Wi}$, alongside the corresponding local rms fluctuations $\delta_{rms} =\sqrt{\bigl\langle(\delta-\bar{\delta})^2\bigr\rangle_t}$  [Fig.~\ref{images}]. For the ordered staggered geometry ($\beta=0$, first row of Fig.~\ref{images}), at low $\text{Wi} \approx 0.75$ each post has an associated downstream wake of high retardation, of similar appearance to that seen downstream of isolated cylinders \cite{Haward2019}. This indicates that the downstream stagnation point of each post is effective at orienting and stretching the micellar microstructure. At $\text{Wi} \approx 0.75$, the flow is steady and the rms fluctuations are low. For increasing $\text{Wi}$, the flow becomes time-dependent and at $\text{Wi} \approx 7.5$ strong fluctuations are observed. The fluctuation appears as the transverse wagging motion of each downstream wake (Movie~1 \cite{ESI}). The growth of fluctuations with $\text{Wi}$ is consistent with the experiments of Walkama \textit{et al.} \cite{Walkama2020} using a similar staggered hexagonal array and range of $\text{Wi}$. We further extend the range of $\text{Wi}$ and observe that fluctuations remain, but apparently become less intense for $\text{Wi}\gtrsim 7.5$. At $\text{Wi}\approx 75$, the fluctuations appear more uniformly throughout the field and are of increased frequency compared with $\text{Wi}\approx 7.5$ (Movie~2 \cite{ESI}).

In the ordered aligned geometry ($\beta=0$, second row of Fig.~\ref{images}), the retardation field at $\text{Wi} \approx 0.75$ is qualitatively different from the staggered array. Here, there is no significant downstream wake and the regions of high stress are concentrated upstream and to the sides of each post. As $\text{Wi}$ is increased, the retardation becomes concentrated in the gaps between the rows of aligned posts. Fluctuations in the ordered aligned array remain low relative to the staggered array until for $\text{Wi}\approx 75$ a similar level of time-dependence is observed throughout the field. Movie~3 and Movie~4 show the time-resolved retardation at $\text{Wi}\approx 7.5$  and $\text{Wi}\approx 75$, respectively \cite{ESI}.

For increasing disorder of the aligned array ($\beta=0.1$ and $\beta=0.3$, third and fourth rows of Fig.~\ref{images}, respectively), at low $\text{Wi} \approx 0.75$ a few posts appear to acquire high stress in downstream wake regions, although unlike the staggered array, these are not necessarily aligned in the primary flow ($x$) direction owing to the disorder. At intermediate $\text{Wi} \approx 7.5$ and $\text{Wi} \approx 12.5$, the flow becomes unsteady, and although less intense than for the staggered array, fluctuations are clearly greater than for the aligned array with $\beta =0$. For the highest $\text{Wi}$ shown ($\text{Wi} \approx 75$), fluctuations are roughly similar in all arrays.

\begin{figure}
    \centering
    \includegraphics[width=8.6cm]{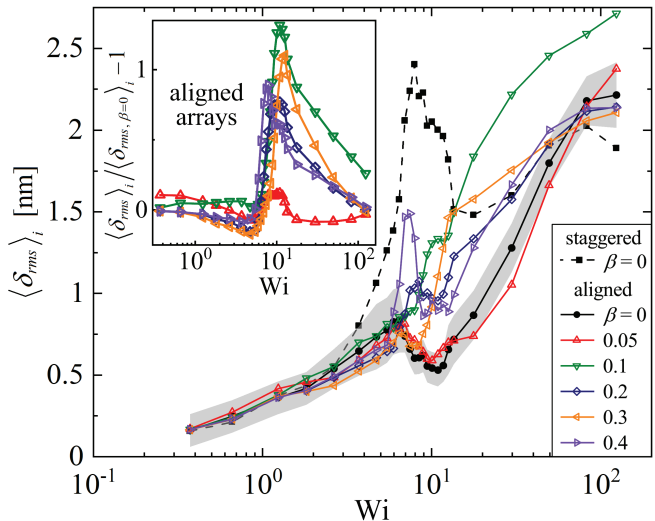}
\vspace{-0.2in}
    \caption{Spatially-averaged retardation fluctuations $\langle\delta_{rms}\rangle_i$ \emph{vs} $\text{Wi}$ in various post arrays. The shaded gray region about the data for the aligned array with $\beta=0$ indicates the typical variability over repeated test runs. Insert compares fluctuations between the ordered and disordered aligned arrays.  
 }
    \label{flucts}
\end{figure}

Spatially-averaged rms retardation fluctuations $\langle \delta_{rms} \rangle_i$ in the staggered array show a local peak at intermediate $\text{Wi}\approx 7.5$, followed by a reduction and a subsequent growth towards an apparent high-$\text{Wi}$ plateau [Fig.~\ref{flucts}]. In the aligned geometry ($\beta=0$), the peak at intermediate $\text{Wi}$ is greatly diminished, although at higher $\text{Wi}$ fluctuations reach a similar plateau value as for the staggered array. With increasing disorder of the aligned geometry, the fluctuations at intermediate $\text{Wi}$ increase. At $\beta=0.4$, a peak emerges at a similar $\text{Wi}$ as the peak seen in the staggered array. At higher $\text{Wi}$, most of the geometries tend to a similar limiting plateau in the level of fluctuation, however at $\beta=0.1$, fluctuations remain higher than for those $\beta=0$.

Spatially-averaged fluctuations from the disordered arrays are compared against those from the ordered aligned array as $(\langle \delta_{rms} \rangle_i - \langle \delta_{rms,\beta=0} \rangle_i)/\langle \delta_{rms,\beta=0} \rangle_i $ [Fig.~\ref{flucts}~(insert)]. Within experimental error, the data for $\beta =0.05$ does not deviate significantly from zero, but for $\beta >0.05$, all of the disordered arrays show a relative peak in the fluctuations around $\text{Wi}\approx 10$ where $\langle \delta_{rms} \rangle_i  \approx 2 \times \langle \delta_{rms,\beta=0} \rangle_i$. Perhaps surprisingly, the increase in fluctuations is saturated (or maximal) for a rather low disorder of $\beta =0.1$. It is noted that Walkama \textit{et al.} showed that small amounts of disorder were also sufficient to strongly \emph{suppress} the fluctuations occuring in a staggered post array configuration  \cite{Walkama2020}. 

\begin{figure}
    \centering
    \includegraphics[width=8.6cm]{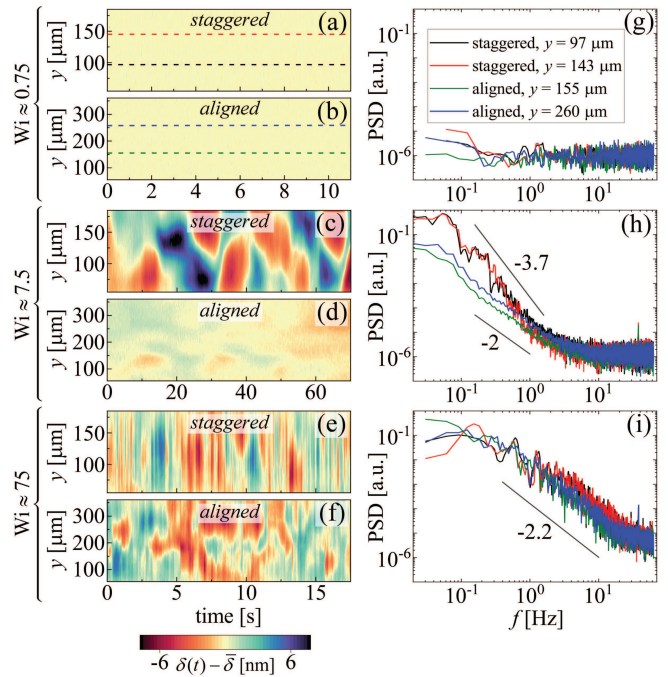}
\vspace{-0.2in}
    \caption{Analysis of retardation fluctuations about the local mean $\delta(t)-\bar{\delta}$ in the ordered staggered and aligned arrays ($\beta=0$). (left) Kymographs of the fluctuations along $x=0$ for: (a,b) $\text{Wi} \approx 0.75$; (c,d) $\text{Wi} \approx 7.5$; (e,f) $\text{Wi} \approx 75$. (right) Power spectral density (PSD) distributions of the fluctuations for: (g) $\text{Wi} \approx 0.75$; (h) $\text{Wi} \approx 7.5$; (i) $\text{Wi} \approx 75$, taken along the correspondingly-colored horizontal dashed lines in (a,b).
}
    \label{S-T}
\vspace{-0.1in}
\end{figure}

Analysis of retardation fluctuations about the local mean $\delta(t)-\bar{\delta}$ in the ordered staggered and aligned arrays ($\beta=0$) at three representative values of $\text{Wi}$ is provided in Fig.~\ref{S-T}.  Kymographs of the fluctuations [Fig.~\ref{S-T}(a-f)] are extracted along the line $x=0$ between the post located at the origin and its neighbor to positive $y$. For low $\text{Wi} \approx 0.75$, the fluctuations in both the staggered [Fig.~\ref{S-T}(a)] and ordered aligned [Fig.~\ref{S-T}(b)] arrays are weak and the kymographs indicate the noise level of the measurement. At this $\text{Wi}$, the power spectral density, PSD, of the fluctuations with time along the four colored dashed lines all show a uniformly flat frequency response [Fig.~\ref{S-T}(g)]. At intermediate $\text{Wi} \approx 7.5$, the strong fluctuations in the staggered array [Fig.~\ref{S-T}(c)] translate to a PSD with high power at low frequencies, that decays steeply into the noise with a power-law exponent $-3.7$  [Fig.~\ref{S-T}(h)]. In contrast, in the ordered aligned array at $\text{Wi} \approx 7.5$ [Fig.~\ref{S-T}(d,h)], the fluctuations are an order-of-magnitude weaker at low frequencies and decay with an exponent $-2$. At high $\text{Wi} \approx 75$, both the staggered and the ordered aligned array show fluctuations of similar intensity [Fig.~\ref{S-T}(e,f)] and with similar frequency content in the PSD [Fig.~\ref{S-T}(i)]. The PSD is shifted to higher frequencies than at $\text{Wi} \approx 7.5$ and decays with an exponent of $\approx -2.2$. The power-law decays in the power spectra of the fluctuating signals indicate that the fluctuations are aperiodic and the slopes are consistent with values reported in studies of elastic turbulence \cite{Qin2017,Walkama2020,Ekanem2020,Browne2020b,Steinberg2021}. 

\begin{figure}
    \centering
    \includegraphics[width=8.6cm]{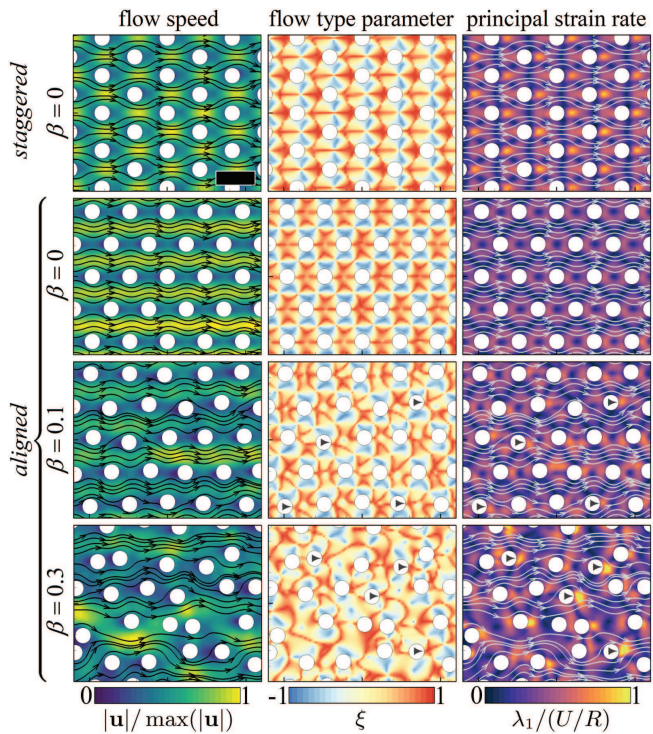}
\vspace{-0.2in}
    \caption{Flow velocimetry in various post arrays at $\textnormal{Wi} \approx 0.75$: (left) Normalized flow speed $\abs{\textbf{u}} /\max(\abs{\textbf{u}})$; (middle) flow type parameter $\xi$; (right) normalized principal strain rate $\lambda_1/(U/R)$. Disordering the aligned array introduces stagnation points, where extension-dominated regions also have high strain rates (examples indicated by gray arrow-heads). Scale bar: $250~\upmu$m.
}
    \label{flowfields}
\vspace{-0.1in}
\end{figure}

Our results reveal a broad consistency with those of Walkama \textit{et al.} \cite{Walkama2020} for viscoelastic flow through an ordered staggered array of posts. However, introducing disorder to an aligned array has a diametrically opposite effect from that shown by Walkama \textit{et al.} for introducing disorder to a staggered array. 
The results can be understood in context with each other by examining the flow fields through the arrays at low $\text{Wi}$ prior to the onset of instability. Flow velocimetry was performed on a FlowMaster volume illumination micro-particle image velocimetry system (LaVision GmbH \cite{Wereley2010}) by seeding the WLM test fluid with $2~\upmu$m diameter fluorescent tracers (PS Fluored, Microparticles GmbH). From the measured velocity fields $\textbf{u} = (u,v)$, the deformation rate  $\textbf{D}$ and vorticity $\pmb{\Omega}$ tensors are computed. Fig.~\ref{flowfields}~(left column), shows the normalized magnitude of the velocity fields $\abs{\textbf{u}} / \max(\abs{\textbf{u}})$. To indicate the local flow kinematics, Fig.~\ref{flowfields}~(middle column) shows the ``flow type parameter" $\xi = (\abs{\bf{D}} - \abs{\pmb{\Omega}} )/( \abs{\bf{D}} + \abs{\pmb{\Omega}})$, where $\abs{\textbf{D}}=\sqrt{2 \textbf{D}:\textbf{D}}$ and $\abs{\pmb{\Omega}}=\sqrt{2 \pmb{\Omega}:\pmb{\Omega}}$ \cite{Astarita1979}. Here, $\xi = -1$ indicates solid body rotation, $\xi = 0$ simple shear, and $\xi = 1$ pure extension. The flow strength in the extensional regions is quantified by the principle strain rate (or eigenvector of $\bf{D}$ \cite{Hamlington2008}),  $\lambda_1 = \frac{1}{2}\sqrt{(\textbf{D}_{11} - \textbf{D}_{22})^2 + 4\textbf{D}_{12}^2}$ [Fig.~\ref{flowfields}~(right column)]. In the staggered array [Fig.~\ref{flowfields}~(top row)], it is evident that streamlines diverge upstream of each post and reconverge downstream, resulting in extension-dominated regions at the up- and downstream stagnation points. Extensional rates are particularly high downstream of each post, where elastic stresses are also high [Fig.~\ref{images}] and time-dependence is first manifested (Movie~1, \cite{ESI}). By contrast, in the ordered aligned array [Fig.~\ref{flowfields}~(second row)] the flow is concentrated in between the aligned rows of posts and streamlines do not connect between successive streamwise-oriented posts. Every stagnation point is thus effectively ``screened" from the flow field (note that although the flow type parameter shows regions of extensional flow between the posts, this is inevitable due to the symmetry of the flow about $y=0$, however $\abs{\lambda_1}$ in these extensional regions is weak). As disorder is applied to the aligned array [Fig.~\ref{flowfields}~(third and fourth rows)], locations emerge where streamlines split and reconverge, reintroducing stagnation points to the flow field. It becomes possible to identify posts downstream of which the flow kinematics are extension dominated and with relatively high streamwise oriented strain rate (examples marked by gray arrowheads in Fig.~\ref{flowfields}). 

Contrary to Walkama \textit{et al.} \cite{Walkama2020}, our results clearly show that disorder does not necessarily suppress chaotic fluctuations in viscoelastic flows. Depending on the initially ordered configuration, disorder can in fact, promote such fluctuations over a range of $\text{Wi}$. In general, this can be understood by considering the rate of occurrence of stagnation points in the flow field. Stagnation points are locations prone to elastic instability due to the high tensile stresses they induce and the consequent feedback on the flow \cite{McKinley1996,Oztekin1997,Groisman2003,Arratia2006,Poole2007,Soulages2009,Haward2012,Haward2016,Sousa2018,Haward2019,Qin2019,Haward2020,Hopkins2021}. In the ordered staggered configuration considered by Walkama \textit{et al.}, the number of such points is maximized (every array element contributes both an upstream and downstream stagnation point). Thus, disordering a staggered array can only serve to reduce the incidence of such locations: some posts will become hidden in the wakes of others, thus screening their stagnation points. By contrast, in the ordered aligned array every stagnation point is screened from the flow field by the previous upstream post (i.e., stagnation points are minimized). In this case, introducing disorder must inevitably increase the occurrence of stagnation points where instability is most probable to initiate. An equivalent, perhaps more intuitive, way to express this is: disordering a staggered array tends to open free paths for the fluid flow (as shown in Ref.~\onlinecite{Walkama2020}), but disordering an aligned array acts to block the free paths that previously existed. Naturally, for high disorders, both staggered and aligned configurations should tend towards randomness, hence homogeneity. We also note that fluctuations become effectively geometry-independent at sufficiently high $\text{Wi}$, where both staggered and aligned configurations show elastic-turbulent-like characteristics.

This work demonstrates the crucial importance of geometry in determining the onset and strength of chaotic fluctuations in viscoelastic flows. In particular, a disorder of just $\beta = 0.1$ applied to an aligned array can increase rms fluctuations by up to $\approx100\%$, highlighting the significant effect that a relatively few stagnation points can have on the global dynamics of viscoelastic flows. 

We gratefully acknowledge the support of the Okinawa Institute of Science and Technology Graduate University (OIST) with subsidy funding from the Cabinet Office, Government of Japan, and also funding from the Japan Society for the Promotion of Science (JSPS, Grant Nos. 18K03958, 18H01135, 20K14656 and 21K03884) and the Joint Research Projects (JRPs) supported by the JSPS and the Swiss National Science Foundation (SNSF). We are grateful to Mr. Kazumi Toda-Peters (OIST) for device fabrication, and to Dr. Stylianos Varchanis (OIST) for helpful discussions.


\providecommand{\noopsort}[1]{}\providecommand{\singleletter}[1]{#1}%

\end{document}